\documentstyle[preprint,aps,tighten,epsf,floats]{revtex}

\newcommand{\nn}{\nonumber\\}
\newcommand{\NN}{$N\!N$}

\newcommand{\NNNN}{$N\!N\!\rightarrow\!N\!N$}
\newcommand{\piN}{$\pi N$}
\newcommand{\piNN}{$\pi N\!N$}

\newcommand{\piNNpiNN}{$\pi N\! N\!\rightarrow\!\pi N\!N$}
\newcommand{\NNtopiNN}{$N\!N\!\rightarrow\!\pi N\!N$}
\newcommand{\piNNtoNN}{$\pi N\!N\!\rightarrow\! N\!N$}
\newcommand{\VOPE}{V_{N\!N}^{\mbox{\protect\scriptsize OPE}}}

\newcommand{\be}{\begin{equation}}
\newcommand{\ee}{\end{equation}}
\newcommand{\bea}{\begin{eqnarray}}
\newcommand{\eea}{\end{eqnarray}}
\newcommand{\eqn}[1]{\label{#1}}
\newcommand{\eq}[1]{Eq.~(\ref{#1})}

\newcommand{\fign}[1]{\label{#1}}
\newcommand{\fig}[1]{Fig.~\ref{#1}}

\newcommand{\bi}{\bibitem}
\newcommand{\F}{{\cal F}}

\newcommand{\V}{{\cal V}}
\newcommand{\G}{{\cal G}}
\newcommand{\T}{{\cal T}}

\newcommand{\bu}{\bar{u}}

\newcommand{\GNN}{D_0}

\newcommand{\TNN}{T_{N\!N}}
\newcommand{\bT}{{\bar{T}}}
\begin{document}

\title{Implementing PCAC in Nonperturbative Models of Pion Production}
\author{B. Blankleider, A. N. Kvinikhidze
\thanks{On leave from the Mathematical Institute of
Georgian Academy of Sciences, Tbilisi, Georgia}}
\address{Department of Physics, Flinders University of South Australia,
Bedford Park, SA 5042, Australia}

\sloppy

\maketitle
\begin{abstract}
Traditional few-body descriptions of pion production use integral equations to
sum the strong interactions nonperturbatively. Although much physics is thereby
included, there has not been a practical way of incorporating the constraints of
chiral symmetry into such approaches.  Thus the traditional few-body
descriptions fail to reflect the underlying theory of strong interactions, QCD,
which is largely chirally symmetric. In addition, the lack of chiral symmetry in
the few-body approaches means that their predictions of pion production are in
principle not consistent with the partial conservation of axial current (PCAC),
a fact that has especially large consequences at low energies. We discuss how
the recent introduction of the ``gauging of equations method'' can be used to
include PCAC into traditional few-body descriptions and thereby solve this long
standing problem
\end{abstract}

\section{Introduction}
Approximate chiral symmetry is an important property of quantum chromodynamics
(QCD) and should therefore be an attribute of any effective description of
strong interaction processes. Yet most nonperturbative effective descriptions of
pion production in few-body processes ($NN\rightarrow\pi NN$, $\pi
N\rightarrow\pi\pi N$, $\gamma N\rightarrow\pi N$, etc.) have been based on
traditional few-body approaches which are not consistent with chiral symmetry --
even if they are based on chiral Lagrangians (i.e.\ the input functions are
chirally invariant). The essential idea of a traditional few-body approach is to
take as input two-body $t$ matrices and pion production vertices, and then use
integral equations to sum the multiple-scattering series nonperturbatively. A
feature of such an approach is that the input can be constructed
phenomenologically without the need to specify an explicit Lagrangian for the
strong interactions. Despite the large amount of physics taken into account
(e.g.\ all possible pair-like interactions are included), the lack of
approximate chiral symmetry in this approach results in a pion production
amplitude that does not obey PCAC.  Although this is a particularly serious
problem at low energies where low-energy theorems apply, the absence of PCAC is
undesirable at any energy because of its inconsistency with QCD. In this
contribution we show how the gauging of equations method \cite{nnn1,nnn2} can be
used to construct a nonperturbative few-body approach that is consistent with
chiral symmetry and whose pion production amplitude obeys PCAC.

\section{Traditional few-body model of $\mathbf{NN\rightarrow\pi NN}$ without
PCAC}


We shall base our discussion on the example of the relativistic \piNN\ system
for which a traditional few-body approach has recently been developed
\cite{KB4d,PA4d}. For simplicity of presentation we shall treat the nucleons as
distinguishable. The few-body approach to the \piNN\ system provides a
nonperturbative simultaneous description of the processes \NNtopiNN, \NNNN, and
\piNNpiNN\ within the context of relativistic quantum field theory.  In this
case the integral equations are four-dimensional and can be expressed
symbolically as\footnote{In this paper it should be understood that all
amplitudes and currents with external nucleon legs are actually operators in
Dirac space and need to be sandwiched between appropriate Dirac spinors
$\bu$ and $u$ to obtain the corresponding physical quantities.}
\be
\T = \V + \V \G_t \T \eqn{BS}
\ee
where, for distinguishable nucleons, $\T$, $\V$, and $\G_t$ are $4\times 4$
matrices written as
\be
\T=\left(\begin{array}{cc} T_{N\!N} & \bT_{N} \\
    T_{N} & T \end{array} \right); \hspace{.3cm}
\V=\left( \begin{array}{cc} V_{N\!N} & \bar{\F}\\
    \F & G_0^{-1}{\cal{I}} \end{array} \right); \hspace{.3cm}
\G_t=\left(\begin{array}{cc} \GNN & 0 \\
    0 & G_0 w^0 G_0 \end{array} \right) . \eqn{AB}
\ee
The elements of matrix $\T$ are defined as follows. $T$ is a $3\times 3$
matrix whose elements $T_{\lambda\mu}$ are Alt-Grassberger-Sandhas (AGS)
amplitudes (generalised to four dimensions) describing the process
\piNNpiNN. Note that the following ``subsystem-spectator'' labelling convention
is used: $\lambda=1$ or $2$ labels the channel where nucleon $\lambda$ forms a
subsystem with the pion, the other nucleon being a spectator, while $\lambda=3$
labels the channel where the two nucleons form the subsystem with the pion being
the spectator.  In a similar way, $T_{N\!N}$ is the amplitude for \NN\
scattering while $T_N$ and $\bT_N$ are $3\times 1$ and $1\times 3$ matrices
whose elements $T_{\lambda N}$ and $T_{N\mu}$ describe \NNtopiNN\ and \piNNtoNN,
respectively. For simplicity of presentation we shall neglect connected diagrams
that are simultaneously \NN - and \piNN - irreducible. Then the elements making
up the kernel matrix $\V$ specified in \eq{AB} take the following form:
\be V_{N\!N}=\VOPE{} - \Delta \eqn{V_NN}
\ee
where $\VOPE{}$ is the nucleon-nucleon one pion exchange potential and
$\Delta$ is a subtraction term that eliminates overcounting.  $\F$ is a
$3\times 1$ matrix with
\be
\F_\lambda =\sum_{i=1}^2\bar{\delta}_{\lambda i}F_i - B \eqn{F}
\ee
where $F_i=f_id_j^{-1}$ consists of $f_i$, the vertex for $N_i\rightarrow \pi
N_i$, and $d_j$, the Feynman propagator of nucleon $j\ne i$. The subtraction
term $B$ in \eq{F} likewise eliminates overcounting.  $\bar{\F}$ is the
$1\times 3$ matrix that is the time reversed version of $\F$, $G_0$ is the
\piNN\ propagator, and ${\cal I}$ is the $3\times 3$ matrix whose
$(\lambda,\mu)$'th element is $\bar{\delta}_{\lambda,\mu}$.  Finally the
propagator term $\G_t$ is a diagonal matrix consisting of the \NN\ propagator
$\GNN$, and the $3\times 3$ diagonal matrix $w^0$ whose diagonal elements are
$t_1 d_2^{-1}$, $t_2 d_1^{-1}$, and $t_3 d_3^{-1}$, with $t_\lambda$ being the
two-body t matrix for the subsystem particles in channel $\lambda$ (for
$\lambda=1$ or $2$, $t_\lambda$ is defined to be the \piN\ $t$ matrix with the
nucleon pole term removed).  The subtraction terms $\Delta$ and $B$ are defined
with the help of \fig{x} as follows:
\be
\Delta=W_{\pi\pi}+W'_{\pi N}+W_{N\!N}+X+Y'-\bar{B}'G_0 B' \ee
where $W'_{\pi N}=W_{\pi N}+PW_{\pi N}P$, $Y'=Y+PYP$, and $B'=B+PBP$, $P$
being the nucleon exchange operator.
\begin{figure}[t]
\centerline{\epsfxsize=11.5cm\epsffile{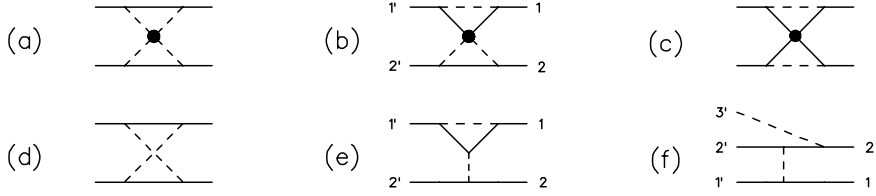}}
\vspace{4mm}

\caption{\fign{x} Diagrams making up the subtraction terms.
(a) $W_{\pi\pi}$, (b) $W_{\pi N}$, (c) $W_{N\!N}$, (d) $X$, (e) $Y$, and (f)
$B$. The dark circles represent the following two-body amplitudes: (a) full
$\pi\pi$ t-matrix, (b) one-nucleon irreducible $\pi N$ t-matrix, and (c) full
\NN\ t-matrix minus the \NN\ one-pion-exchange potential.}
\end{figure}

Despite the rich amount of physics incorporated into the \NNtopiNN\ amplitude
within this model, it becomes evident from the following discussion that
this amplitude cannot satisfy PCAC.

\section{New few-body model of $\mathbf{NN\rightarrow\pi NN}$ with PCAC}

\subsection{Choosing appropriate degrees of freedom}

Although we do not specify the exact Lagrangian behind our new few-body
approach, we do assume that this underlying Lagrangian is chirally invariant (up
to a small explicit chiral symmetry breaking term). In turn, the chiral
invariance of the Lagrangian puts a strong constraint on the nature of the
fields that can be used to construct a practical few-body model. For example, if
one would like to follow the \piNN\ model above and have only pions and nucleons
as the degrees of freedom, then the underlying chiral Lagrangian would
necessarily involve an isovector pion field $\vec{\xi}$ that transforms into
functions of itself under a chiral transformation. Unfortunately, this can
happen only in a nonlinear way \cite{Weinberg}:
\be
\vec{\xi} \rightarrow \vec{\xi} + \frac{1}{2}\vec{\theta}(1-\vec{\xi}^2)
+\vec{\xi} \vec{\theta}\cdot\vec{\xi}
\ee
where $\vec{\theta}$ is the vector of three (infinitesimally small) rotation
angles. Thus one pion transforms into two pions, two pions transform into four
pions, etc., a situation that would make it difficult to formulate a chirally
invariant few-body description (i.e.\ a description whose exposed states
have a restricted number of pions -- in our case 0 or 1). Alternatively,
we can follow the example of the linear sigma model and choose an underlying
Lagrangian that involves, in addition to pions and nucleons, the isoscalar field
$\sigma$. In this case the $\sigma$, together with the pion field $\vec{\phi}$,
transform under the chiral transformation in a linear way:
\be
\vec{\phi}\rightarrow\vec{\phi}+\vec{\theta}\sigma,\hspace{1cm}
\sigma\rightarrow \sigma - \vec{\theta}\cdot\vec{\phi}.
\ee
Thus the four-component field $\phi\equiv(\sigma,\vec{\phi})$ transforms into
itself - a situation that is ideal for a few-body approach. We shall therefore
adopt the latter approach and treat pions and sigma particles on an equal
footing.  For the few-body \piNN\ model of Sec.\ 2, this means mostly a formal
change where the usual (isospin) three-component pion is replaced by a
four-component one. There is, however, one new aspect in that terms with a
three-meson vertex ($\pi\pi\sigma$) now need to be included. For example, one
will need the new subtraction term illustrated in \fig{sub}. Nevertheless, the
equations of Sec.\ 2 retain their structure, the only essential change being an
increase in the size of the matrices like $T$, $T_N$ and $\bT_N$ to take into
account the introduction of a $\sigma N\!N$ channel. With these modifications
we obtain few-body \piNN\ equations that are
consistent with an underlying chiral Lagrangian.
\begin{figure}[h]
\centerline{\epsfxsize=1.4cm\epsffile{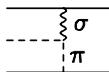}}
\vspace{4mm}

\caption{\fign{sub} New subtraction term in the few-body \piNN\ model where
pions and sigmas are treated on an equal footing.}
\end{figure}
\subsection{Coupling the axial vector field everywhere}

Now that we have achieved consistency with an underlying strong interaction
chiral Lagrangian, our next goal is to construct a few-body \piNN\ model whose
axial current is partially conserved. The problem at hand is analogous to the
one of constructing a conserved electromagnetic current for a few-body system
whose strong interactions are consistent with an underlying Lagrangian that
conserves charge.  In this case it is well known that exact current conservation
is obtained by coupling an external electromagnetic field to all possible places
in the strong interaction model. In a similar way, one would expect to obtain a
partially conserved axial current by coupling an external axial vector field to
all possible places in the strong interaction model. Until recently, however,
what has not been known is the way to achieve this complete coupling for a
few-body system whose strong interactions are described nonperturbatively by
integral equations. Fortunately, the solution to this problem presented recently
in the context of electromagnetic interactions \cite{nnn1,nnn2}, is based on a
topological argument and therefore applies equally well to the present case of
an axial vector field.

The basic idea is to add a vector index (indicating an external axial vector
field) to all possible terms in the {\em integral equations} describing the
strong interaction model. Thus, in the case of the \piNN\ system where the
structure of the integral equations is given as in \eq{BS}, coupling an external
axial isovector field gives
\be 
\T^\mu = \V^\mu + \V^\mu \G_t \T + \V \G_t^\mu \T + \V \G_t \T^\mu
\ee
which can easily be solved to give a closed expression for $\T^\mu$:
\be
\T^\mu = (1+\T\G_t)\V^\mu(1+\G_t\T) + \T \G_t^\mu \T \eqn{T^mu}.
\ee
$\T^\mu$ is a matrix of transition amplitudes $T^\mu_{N\!N}$, $T^\mu_{N\Delta}$,
$T^\mu_{Nd}$, etc.\ (for the moment, we suppress the isovector index in these
amplitudes).  Here we are particularly interested in the transition amplitude
$T^\mu_{N\!N}$ as it is closely related to the pion production amplitude we
seek.

It is easy to see that the transition amplitude $T^\mu_{N\!N}$ given by
\eq{T^mu} has the axial vector field attached everywhere in $T_{N\!N}$ {\em
except} on the external nucleon legs \cite{nnn1}. Including these external leg
contributions then gives the complete axial vector transition current
for \NNNN:
\be
j^\mu_{N\!N} = (\Gamma_1^\mu d_1+\Gamma_2^\mu d_2) \TNN
+ \TNN(d_1\Gamma_1^\mu + d_2\Gamma_2^\mu ) + \TNN^\mu \eqn{j_NN^mu}  \eqn{j^mu}
\ee
where $\Gamma_i^\mu$ is the axial vertex function of nucleon $i$.  The input to
the \piNN\ equations consists of the two-body $t$ matrices $t_i$, the pion
production vertices $f_{i}$, and the single particle propagators $d_{i}$.  Thus
the input to \eq{T^mu} also includes the axial transition currents
$t_i^\mu$, $f_i^\mu$, and axial vertex function $\Gamma_i^\mu$. In order for
these input quantities to be consistent with an external axial vector field
being attached everywhere, they must be constructed to satisfy the Axial
Ward-Takahashi (AWT) identities \cite{Bentz}. This is easily achieved by
restricting the form of the corresponding bare quantities \cite{nnn1}. Thus,
for example, $\Gamma_i^{a\mu}$ needs to satisfy the AWT identity
\be
q_\mu \Gamma_i^{a\mu}(k,p)=i\left[d_i^{-1}(k)\gamma_5 + \gamma_5
d_i^{-1}(p)\right] t^a
- if_\pi m_\pi^2 f_i^a(k,p)d_\pi(q)
\ee
where $p$ and $k$ are the initial and final momenta of the nucleon, $q=k-p$,
$f_i^a(k,p)$ is the \piNN\ vertex for a pion of isospin component $a$, $t^a$ is
an isospin $1/2$ matrix, $m_\pi$ is the mass of the pion, and $f_\pi$ is the
pion decay constant. With the other input quantities constructed to satisfy
similar AWT identities, it can be shown that the two-nucleon axial current given
by \eq{j^mu} satisfies the AWT corresponding to exact PCAC:
\bea
\lefteqn{q_\mu j^{a\mu}_{N\!N}(k_1k_2,p_1p_2)}\nn
&=&i\left[(\gamma_5 t^a)_1 T_{N\!N}(k_1-q,k_2;p_1p_2)+
T_{N\!N}(k_1k_2;p_1+q,p_2)(\gamma_5 t^a)_1\right]\nn
&+&i\left[(\gamma_5 t^a)_2 T_{N\!N}(k_1,k_2-q;p_1p_2)+
T_{N\!N}(k_1k_2;p_1,p_2+q)(\gamma_5 t^a)_2\right]\nn
&-&if_\pi m_\pi^2 T_{N0}^a(k_1k_2,p_1p_2)d_\pi(q)  \eqn{AWT2}
\eea
where $T_{N0}^a$ is the amplitude for $\pi N\!N\rightarrow N\!N$ and in its
time reversed form, is just the the pion production amplitude that we are
seeking. Note that the axial current $j^{a\mu}_{N\!N}$ contains a pion pole
\cite{Weinberg,Bentz}:
\be
j^{a\mu}_{N\!N}(k_1k_2,p_1p_2)= \bar{j}^{a\mu}_{N\!N}(k_1k_2,p_1p_2)
+d_\pi(q)F_\pi^\mu(q) T_{N0}^a(k_1k_2,p_1p_2),
\ee
where $\bar{j}^{a\mu}_{N\!N}$ has no pion pole and $F_\pi^\mu$ is the pion decay
vertex function. Thus an alternative way of obtaining the pion production
amplitude $T_{N0}$, is to take the residue of \eq{j^mu} at the pion pole.

\subsection{Physical content of the new model}

Not only does the pion production amplitude $T_{N0}$ obey exact PCAC, but it
contains a very rich amount of physics that goes beyond what is included in the
traditional few-body approach of Sec.\ 2. A few of the infinite number of new
contributions are illustrated in \fig{new}. 
\begin{figure}[h]
\centerline{\epsfxsize=11.9cm\epsffile{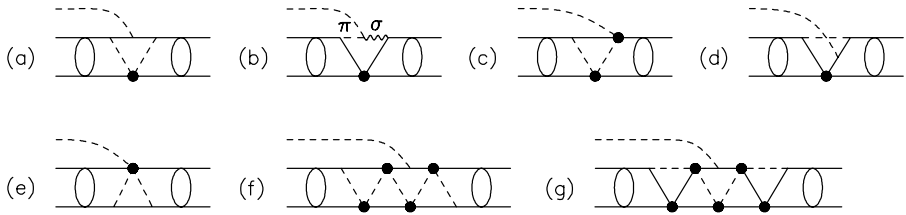}}
\vspace{4mm}

\caption{\fign{new} Examples of new pion production mechanisms included in the
few-body model with PCAC.}
\end{figure}

Finally we would like to stress the practicality of our approach: PCAC and
inclusion of an infinite number of new pion production mechanisms has been
achieved through a simple closed expression, \eq{T^mu}, involving scattering
amplitudes ${\cal T}$ obtained from  a traditional few-body model,
\eq{BS}. Our method also does not depend on the model used for kernel ${\cal V}$
-- rather than \eq{AB}, we could equally have used the simpler case of an \NN\
one-pion-exchange potential.

{\em Acknowledgement}. This work was supported by a grant from the Flinders
University Research Committee.


\begin{thebibliography}{99}
\bi{nnn1} A.\ Kvinikhidze and B.\ Blankleider, Phys.  Rev. C {\bf 60}, 044003
(1999).
\bi{nnn2} A.\ Kvinikhidze and B.\ Blankleider,  Phys.  Rev. C {\bf 60}, 044004
(1999).
\bibitem{KB4d} A.\ N.\ Kvinikhidze and B.\ Blankleider, Nucl.\ Phys.\ {\bf
A574}, 788 (1994).
\bibitem{PA4d} D. R. Phillips and I. R. Afnan, Ann. Phys. (N.Y.) {\bf 247},
19 (1996).
\bi{Weinberg} S. Weinberg, {\em The Quantum Theory of Fields II}
(Cambridge University Press, Cambridge, 1966). 
\bi{Bentz}  W. Bentz, Nucl. Phys. {\bf A446}, 678 (1985).
\end{thebibliography}
\end{document}